\documentclass{IOS-Book-Article}

\usepackage{mathptmx}

\def\hb{\hbox to 10.7 cm{}}

\usepackage{graphicx}

\newcommand{\PMCMC}{\texttt{SPUX}\ }
\newcommand{\figsuffix}{2000p-100s-200c}

\newcommand{\Pobs}{P^\textnormal{\footnotesize{obs}}}
\newcommand{\bx}{\bf{x}}
\newcommand{\bh}{\bf{h}}
\newcommand{\bth}{\bf{\theta}}

\begin{document}

\pagestyle{headings}
\def\thepage{}

\begin{frontmatter}              

\title{SPUX: Scalable Particle Markov Chain Monte Carlo
	for uncertainty quantification in
	stochastic ecological models}

\markboth{}{September 2017\hb}

\author[A]{\fnms{Jonas} \snm{{\v S}ukys}%
\thanks{Corresponding Author: Jonas {\v S}ukys,
	Eawag: Swiss Federal Institute of Aquatic Science and Technology,
	\"Uberlandstrasse 133, CH-8600 D\"ubendorf, Canton of Zurich, Switzerland; E-mail: jonas.sukys@eawag.ch.}}
and
\author[B]{\fnms{Mira} \snm{Kattwinkel}}

\runningauthor{J.{\v S}ukys et al.}
\address[A]{Eawag: Swiss Federal Institute of Aquatic Science and Technology, Switzerland}
\address[B]{University of Koblenz-Landau, Germany}

\begin{abstract}
Calibration of individual based models (IBMs), successful in modeling complex ecological dynamical systems, is often performed only ad-hoc. Bayesian inference can be used for both parameter estimation and uncertainty quantification, but its successful application to realistic scenarios has been hindered by the complex stochastic nature of IBMs. Computationally expensive techniques such as Particle Filter (PF) provide marginal likelihood estimates, where multiple model simulations (particles) are required to get a sample from the state distribution conditional on the observed data. Particle ensembles are re-sampled at each data observation time, requiring particle destruction and replication, which lead to an increase in algorithmic complexity. We present SPUX, a Python implementation of parallel Particle Markov Chain Monte Carlo (PMCMC) algorithm, which mitigates high computational costs by distributing particles over multiple computational units. Adaptive load re-balancing techniques are used to mitigate computational work imbalances introduced by re-sampling. Framework performance is investigated and significant speed-ups are observed for a simple predator-prey IBM model. 
\end{abstract}

\end{frontmatter}
\markboth{September 2017\hb}{September 2017\hb}

\section{Introduction}

\begin{figure}
	\centering
	\includegraphics[width=\textwidth]{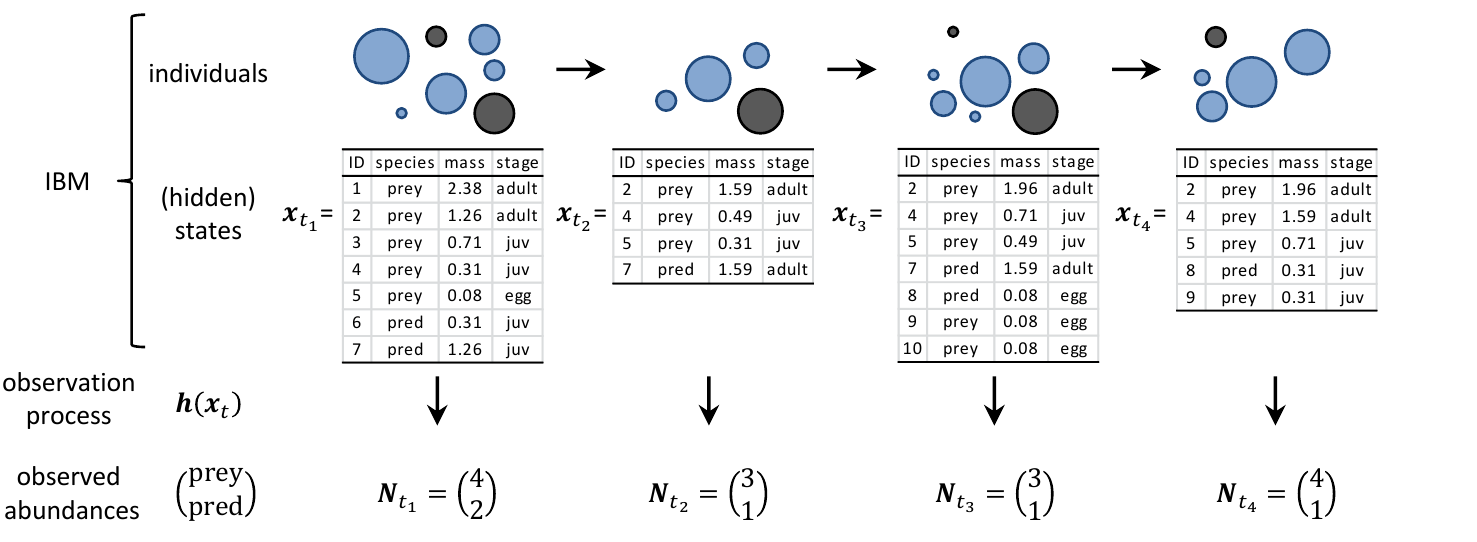}
	\caption{Illustration of a two-species IBM with individuals of either prey or predator species at different life stages and with different masses as a state-space or hidden Markov model. Circles depict individuals in the individual-based simulation (preys are represented in blue, predators in grey), the states $\mathbf{x}_t$ are represented by all properties of the individuals necessary for the evolution to the next time step (only examples of the properties are given here), the incomplete observation process $\mathbf{h}(\mathbf{x}_t)$ filters for detectable (sufficiently large) individuals.
	{\tiny [Reprinted from Environmental Modelling \& Software, 87, Kattwinkel M \& Reichert P, Bayesian parameter inference for Individual-Based Models using a Particle Markov Chain Monte Carlo method, 110--119., Copyright (2017), with permission from Elsevier.]}}
	\label{f:ibm}
\end{figure}

Individual or agent based models (IBMs/ABMs) have been successfully applied for modeling complex ecological dynamical systems \cite{DeAngelis2005,Brown2004}. They simulate all individuals in a population or other group of interest as individual entities and consider the stochastic nature of processes such as birth, death, movement or decision making. Furthermore, IBMs explicitly consider individual characteristics (e.g. mass, life stage, personal preferences) that differ among the individuals, and they can easily incorporate the spatio-temporal arrangement of organisms, and interactions among individuals \cite{DeAngelis2005}. A typical IBM used in this case study is illustrated in Figure \ref{f:ibm}, and consists of individuals of a predator species and a prey species at different life stages and with different masses.

However, model parameter estimation for IBMs is often performed only ad-hoc
by manual tuning or optimization, lacking a thorough assessment of the underlying uncertainties. Bayesian inference \cite{Ellison2004} can be used for both parameter estimation and uncertainty quantification. It is based on the so-called likelihood function $L(D|\theta, M)$ of the model $M$ that formulates the model as a probability distribution of observations $D$ for given model parameter values $\theta$. Additionally, prior information about parameter values is described probabilistically by prior distribution $\pi(\theta)$. This prior knowledge on model structure and parameters is then combined with information from observed data $D$ to the posterior distribution $p(\theta|D, M)$ of model parameters $\theta$:
\begin{equation}
\label{eq:bayes}
p(\theta|D, M) \propto L(D|\theta, M) \pi(\theta).
\end{equation}

\begin{figure}
	\centering
	\includegraphics[width=\textwidth]{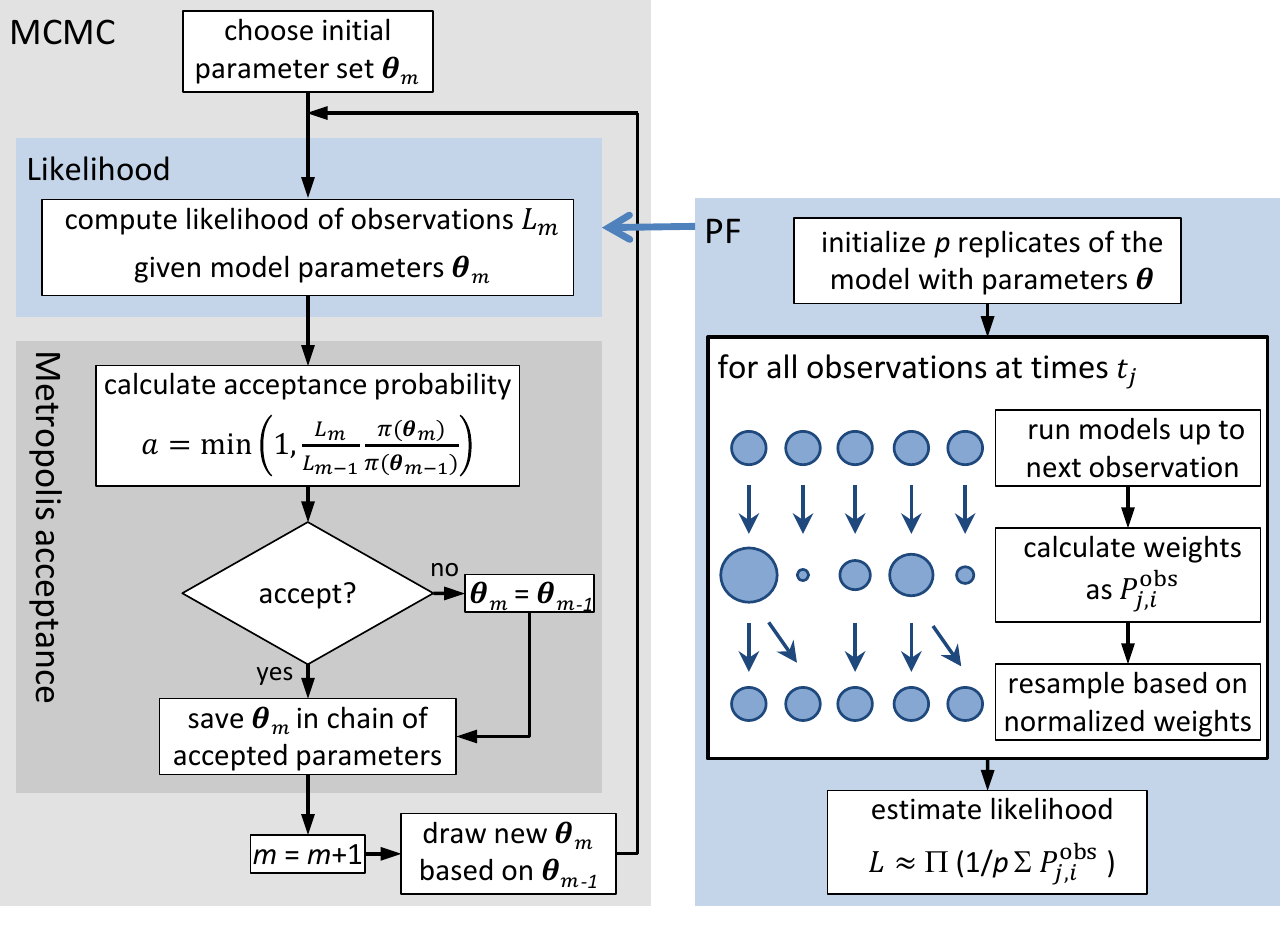}
	\caption{Simplified flowcharts of the PMCMC algorithm combining MCMC (left) for posterior sampling and PF (right) for marginal likelihood estimation in stochastic models using ensembles of re-sampled model states.
	{\tiny [PF reprinted from Environmental Modelling \& Software, 87, Kattwinkel M \& Reichert P, Bayesian parameter inference for Individual-Based Models using a Particle Markov Chain Monte Carlo method, 110--119., Copyright (2017), with permission from Elsevier; altered.]}}
	\label{f:pmcmc}
\end{figure}

Usually, Bayesian inference cannot be solved analytically for posterior $p(\theta|D, M)$. Therefore, numerical schemes have been developed to sample from the posterior distribution of model parameters. Commonly applied techniques are Metropolis or Metropolis-Hastings Markov chain Monte Carlo (MCMC) sampling schemes \cite{Gamerman2006,Gelman2014,Hastings1970,Metropolis1953}. These schemes are based on simulations for a large number of parameter samples and the evaluation of the likelihood and prior at these parameter values.
As presented by a flowchart in Figure \ref{f:pmcmc} (left), parameter values are sampled in a chain: based on the evaluated likelihood for a given set of proposal parameters $\theta$, parameters are either accepted (with a probability based on the ratio of current and previous posterior estimates) or rejected (in which case previously accepted parameters are appended to the chain instead).

The successful application of Bayesian inference for IBMs to realistic scenarios has been hindered by the complex stochastic nature of such models leading to (i) technical difficulties in likelihood estimation and (ii) entailing high computational costs. Likelihood estimation issues were recently tackled by applying Particle Filter (PF) to MCMC sampling, known as Particle Markov Chain Monte Carlo (PMCMC) technique \cite{Andrieu2010}. In PMCMC, the hidden Markov structure of IMBs is exploited for efficient marginal likelihood approximations using time-series observations \cite{Kattwinkel2017}. A model has a Markov structure if the future model states do not depend on the past other than through the current state. The Markov structure is hidden if the observations do not follow a Markov structure, hence, the states are observed incompletely. This is typically the case in IBMs with observations consisting of time-series data with $n$ measurements (e.g. of abundances). The hidden Markov structure is evident for IBM illustrated in Figure \ref{f:ibm} where observations are only available for the abundances of individuals exceeding a certain mass.
Using Markov structure, the marginal likelihood can be rewritten as follows \cite{Kattwinkel2017},
\begin{equation}
\label{eq:L-markov}
L (D | \bth, M) = \int
f_0 \left(\bx_{t_0}|\bth\right)
\prod_{j=1}^n f_j \left(\bx_{t_j}|\bx_{t_{j-1}}, \bth\right)
\prod_{j=1}^n \Pobs (D|\bh(\bx_{t_j}), \bth)
d\bx_{t_0} \dots d\bx_{t_n}.
\end{equation}
Here, for given parameters $\bth$, probability distributions $f_j (\bx_{t_j}|\bx_{t_{j-1}}, \bth)$ characterize random model state vector $\bx_{t_j}$ given previous state $\bx_{t_{j-1}}$, and observation likelihood $\Pobs_j = \Pobs (D|\bh(\bx_{t_j}), \bth)$ is dependent on a specific model and data type \cite{Kattwinkel2017}.
PMCMC algorithm \cite{Andrieu2010,Kattwinkel2017} provides an unbiased statistical estimate of the proposal parameter marginal likelihood $L (D | \bth, M)$ with structure given in equation \ref{eq:L-markov}.
As depicted in Figure \ref{f:pmcmc} (right), marginal likelihood approximation procedure involves multiple evaluations $i = 1, \dots, p$ of the underlying model, so called particles. At each measurement time $t_j$ in the observations time series, model simulations are paused and all particles are re-sampled according to their normalized marginal likelihoods $\Pobs_{j,i}$.
Such periodic re-sampling, however, increases algorithmic complexity due to required destruction and replication of existing particles.
At the end of the Particle Filtering,
an unbiased estimate $\hat L(D | \bth, M)$ of marginal likelihood $L(D | \bth, M)$ as in equation \ref{eq:L-markov} is evaluated by
\begin{equation}
\label{eq:L-PF}
L (D | \bth, M) \approx
\hat L (D | \bth, M) =
\prod_{j=1}^n
\left(\frac{1}{p}\sum_{i=1}^p \Pobs_{j,i} \right).
\end{equation}
PMCMC was already successfully applied to simple IBM \cite{Kattwinkel2017}.
However, complex IBMs modeling realistic ecological systems were out of reach due to limitations of serial particle filtering implementation.

The aim of this manuscript is to address the computational issues in PMCMC using modern high performance computing techniques and to enable Bayesian inference for realistic ecological models. We present \texttt{SPMC}, a Python-based framework for parallel PMCMC, which mitigates high computational costs by adaptively distributing particles (model evaluations) over multiple computational units in a parallel compute cluster.

\section{Parallelization}
\label{s:parallelization}

In particle filter marginal likelihood estimator (see Figure \ref{f:pmcmc} (right)),
each particle in the ensemble is assumed to posses an execution control interface.
Such interface is assumed to expose main model workflow routines,
such as \texttt{init($\bth$)}, \texttt{run($t_j$,seed)}, and \texttt{observe()}.
This ensures efficient execution, since model state is kept in working memory
at the end of every observation time $t_j$, with next simulation phase for time $t_{j+1}$ executed without additional overhead.
During particle re-sampling stage, depending on the acquired observation likelihoods $\Pobs_{j,i}$, excessive replications of strongly-weighted particles would exert a strong imbalance in parallel usage of resources. 
Auxiliary model interface routines \texttt{state=save()} and \texttt{load(state)} are assumed to be provided for writing and reading a serialized model state to and from memory (binary stream), respectively, ensuring efficient particle replication (and inter-node communication, if needed).

\subsection{Paradigm}

\begin{figure}
	\centering
	\includegraphics[width=\textwidth]{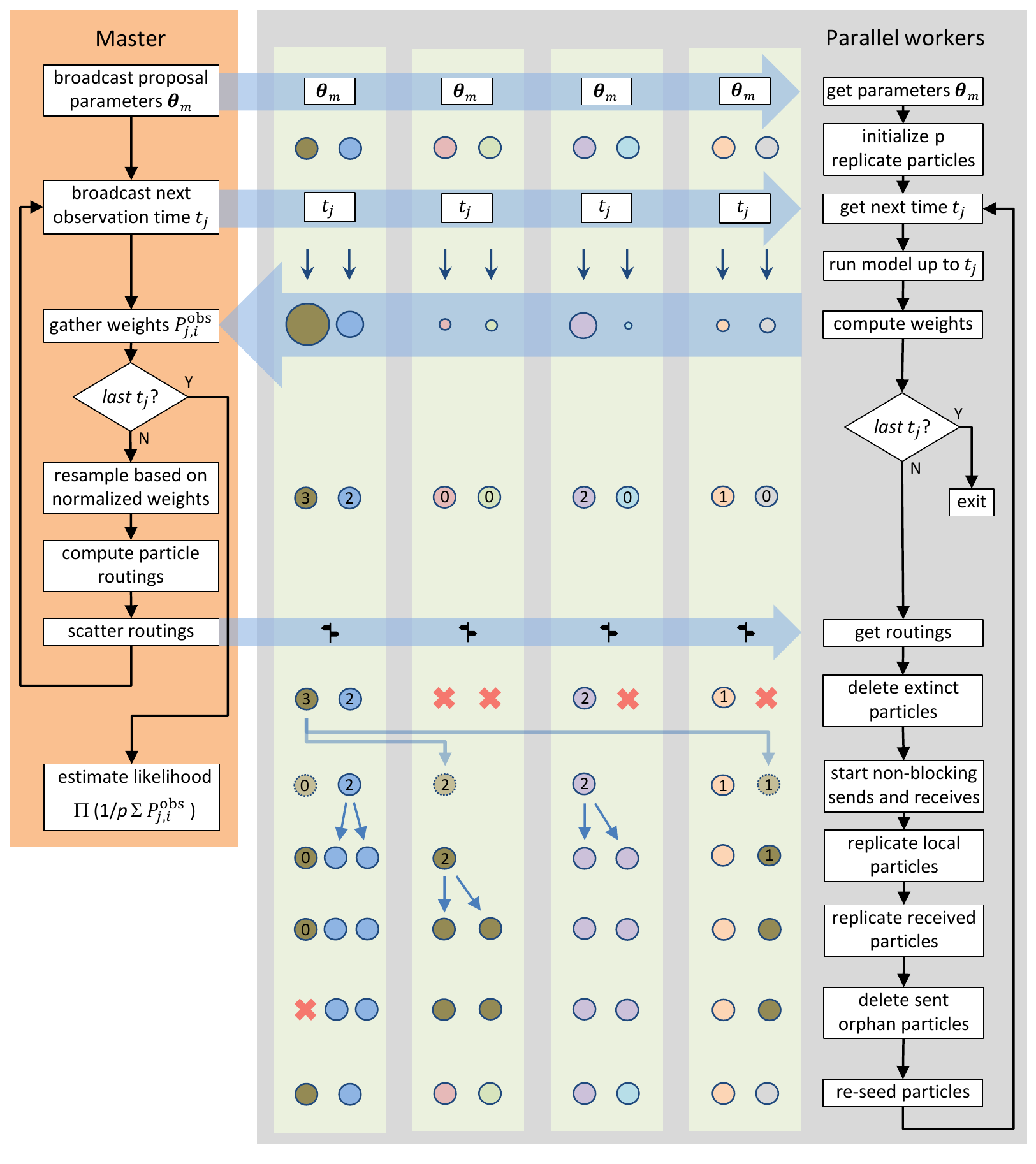}
	\caption{Flowchart of the parallel particle filtering algorithm with adaptive load balancing.
		Requests for multiple model instances (particles) are dispatched to parallel workers, where models are initialized and executed.
		During re-sampling steps, particles are re-distributed from the over-utilized to neighboring under-utilized workers according to routings dispatched from master. Communication and computation overlap is achieved by asynchronous send and receive requests for particle states while performing local particle replications.
	}
	\label{f:ppf}
\end{figure}

Nested parallelization framework is used, with outer and inner parallelizations corresponding to MCMC chains and particle filtering stages of the algorithm, respectively.
Outer parallelization for standard MCMC algorithm is trivial as multiple (often in the order of tens or hundreds) independent MCMC chains can be evaluated in parallel without any required communication (for a standard MCMC sampler).
We focus on the inner parallelization, where multiple parallel workers are executing a partial ensemble of all required particles, with periodic particle filtering (re-sampling)
at specified observation times $t_j$.
A master - worker framework is proposed, where particles are distributed across parallel workers controlled by the master.
Master process would dispatch parameters and observation times to worker processes,
which would in return report back estimated observation likelihoods.
Both types of communication patterns,
master – worker (inter-communicator)
and
worker – worker (intra-communicator),
are required: the latter is relevant for particle routing in order to adaptively re-balance
worker loads after particle re-sampling stages.
A detailed example of parallelization structure, execution and communication patterns
for $p=8$ particles and 4 parallel workers
is presented in Figure \ref{f:ppf}.
Main stages in the parallel particle filtering algorithm are the following:
\begin{enumerate}
	\item Master broadcasts model parameters $\bth$ to all workers.
	\item Each worker initializes an approximately equal fraction of required $p$ particles to obtained parameters $\bth$ by calling \texttt{init(seed)} with globally different seeds for pseudo random number generation in each stochastic model.
	\item Master broadcasts next required observation time $t_j$ to workers.
	\item Workers evolve all particles until time $t_j$ by calling \texttt{run(seed)} routine.
	\item Workers compute observation likelihoods $\Pobs_{j,i}$ for all particles $i = 1, \dots, p$, which are then gathered in the master.
	\item If all required observation times have been processed, master broadcasts exit command to workers (not included in the example for clarity), and evaluates amrginal likelihood estimate $\hat L (D|\bth,M)$ as in equation \ref{eq:L-PF}.
	\item Based on gathered and normalized observation likelihoods $\Pobs_{j,i}$ as probabilistic weights, master re-samples current particle ensemble (virtually, no actual particle instances resident in worker memory are affected).
	\item Master computes routings (refer to following subsection \ref{ss:routings} for details) which contain instructions on adaptive re-distribution of re-sampled particles in order to equilibrate computational work across all workers.
	\item Master scatters relevant part of routings to each worker and jumps to step 3.
	\item Communication and computation overlapping is achieved by employing asynchronous send and receive requests for particle states while performing required local particle replications. In particular, based on information in received routing, each worker performs the following tasks:
	  \begin {enumerate}
	    \item Remove particles that are no longer needed on this worker and do not need to be send to any other worker.
	    \item Initiate asynchronous non-blocking sends for serialized particle states (from \texttt{stat=save()}) which are routed from current worker to other workers and asynchronous non-blocking receives for serialized particle states which are routed from other workers to current worker.
	    \item Replicate (employing \texttt{state=save()} and \texttt{load(state)}) local particles that are routed to current worker itself and more than one particle is required.
	    \item Once asynchronous receives are completed, replicate received particles if more than one particle is required.
	    \item Once asynchronous sends are completed, remove particles that were sent to other workers, but are not needed on this worker.
	    \item Reset seeds for pseudo random number generators (such that replicated particles continue as independent stachastic realizations) and jump to step 3.
	  \end{enumerate}
\end{enumerate}

To further reduce communication overhead, identical particles are moved together
by moving only a single particle and then replicating it on the destination worker.

\subsection{Routings}
\label{ss:routings}

During routing, re-sampled particles are instructed to be moved from over-utilized workers to the neighboring under-utilized workers across their intra-communicator.
Routing objects include worker-specific information regarding particle identification, source (present worker address), destination (re-routed worker address), and re-identification (for post-routing re-seeding).
An empirical greedy algorithm is employed for constructing routings based on re-sampled particle counts and current particle distribution across workers.
The maximal allowed worker load $W_{\max}$ is estimated by dividing the total number of particles by the total number of workers.
Two main stages of the algorithm are:
\begin{enumerate}
	\item \textbf{Prioritize local particle replications:} for each worker, resident re-sampled particles are kept on original workers provided their loads do not exceed $W_{\max}$.
	\item \textbf{Route particles to closest worker:} iterate across remaining particles and route each to a closest (measured by proximity of worker ranks) under-utilized worker.
\end{enumerate}
Such load balancing strategy does not take into account possibly heterogeneous model runtimes for each particle, and hence there could be potential gains from dynamic balancing frameworks using task stealing instead of a static routing approach.

\subsection{Implementation}
\label{ss:implementation}

We have implemented a new parallel PMCMC package \PMCMC as a modular Python framework for Bayesian uncertainty quantification.
Python is a modern high-level programming language and has recently seen increasing interest from the high performance computing community \cite{Smith2016}.
Modular design in \PMCMC allows for simple framework extension by
general hidden Markov models in any programming language and various MCMC posterior samplers
(currently: Metropolis-Hastings \cite{Hastings1970} and EMCEE \cite{Foreman-Mackey2013})
are supported as plug-ins.
User-defined likelihoods and load balancers can be also implemented.

\PMCMC is parallelized using Message Passing Interface (MPI) library \cite{MPI3} via the object-oriented Python MPI bindings package \texttt{mpi4py} \cite{Dalcin2011}.
Pickle-based communication of arbitrary serializable Python objects is used for execution workflow management across master and workers, as the memory and processing time overheads are negligible.
Efficient communication of \texttt{NumPy} arrays (using exposed C-style data buffer)
is used for communication of saved particle states, assumed to be arrays of raw bytes.
MPI dynamic process management for parallel workers is employed for deploying nested parallelization: parallel MCMC chains execute marginal likelihood estimation using parallel particle filtering.
Particle sandboxing prevents race conditions in model implementations.

\section{Applications}
\label{s:applications}

We apply \PMCMC to a simple test case of predator-prey IBM
described in detail in \cite{Kattwinkel2017}.
The stochastic IBM is implemented in Java programming language \cite{Kattwinkel2017}.
In \PMCMC Java classes can be directly accessed in Python runtime using \texttt{jpype} interface \cite{SteveMenard}, avoiding heavy file system access and significantly reducing the overhead of storing or loading model states during re-sampling stages.
During the first 1901 simulations steps, IBM simulations are initialized to approximately 
2000 preys and 30 predators.
Afterwards, the model is executed for 20 available observation times,
until its last step at 2604.
Observation likelihood $\Pobs$ is evaluated based on the observation (counting) process error model provided in \cite{Kattwinkel2017}.
Two calibration parameters were selected, prey and predator half saturation constants for self inhibition, with respective starting values of 25 and 15.
2000 particles were used in Particle Filter for the estimation of marginal parameters likelihood.
With the focus on the performance of the PF rather than of MCMC, only 100 MCMC samples were computed.
There is no ``burn-in'' stage in the MCMC sampling,
since the selected starting values were also used for the generation of synthetic observation data.
\PMCMC was executed on 200 parallel worker cores on EULER cluster in ETH Zurich, Switzerland.

\begin{figure}
	\centering
	\includegraphics[height=0.22\textheight]{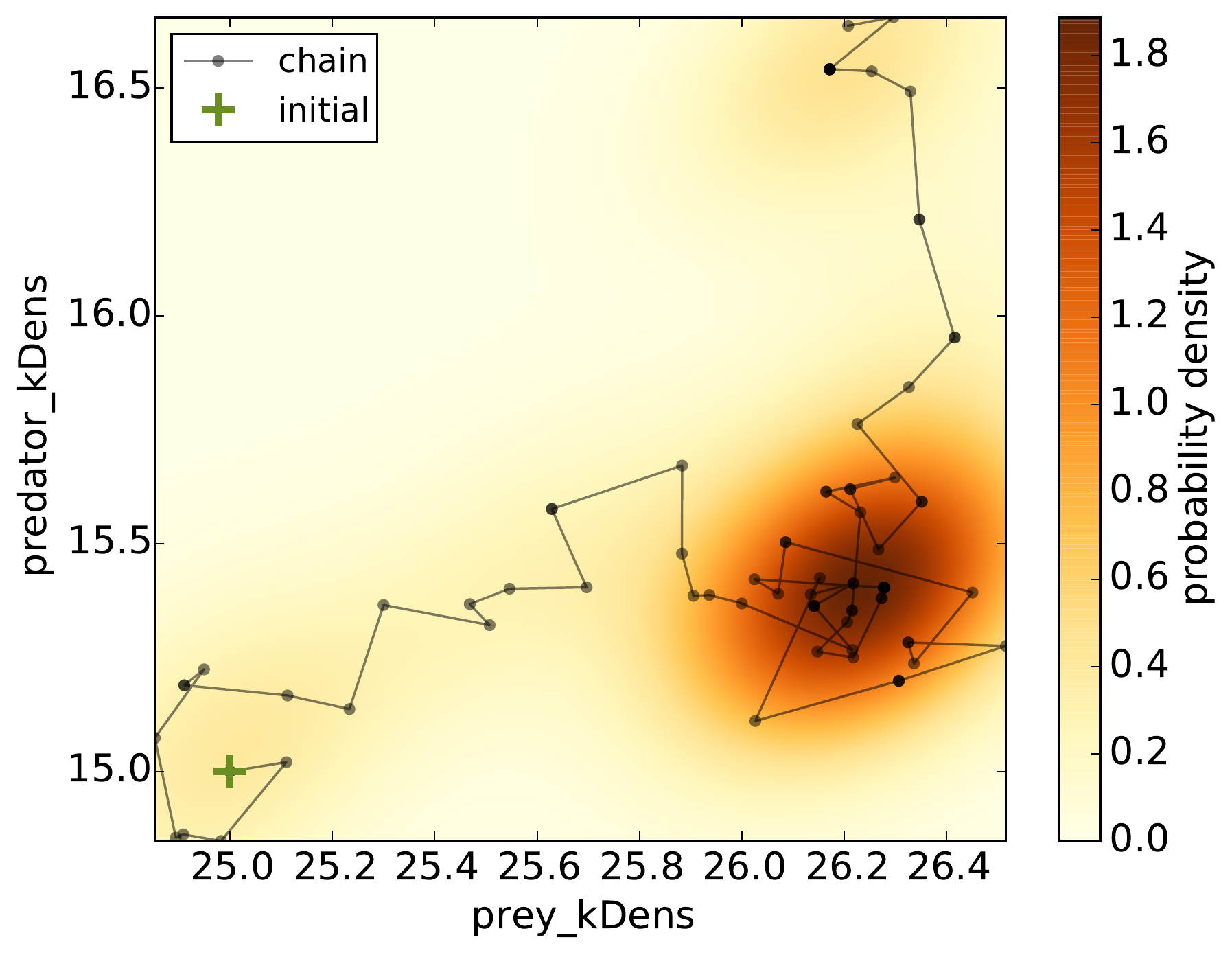}
	\includegraphics[height=0.225\textheight]{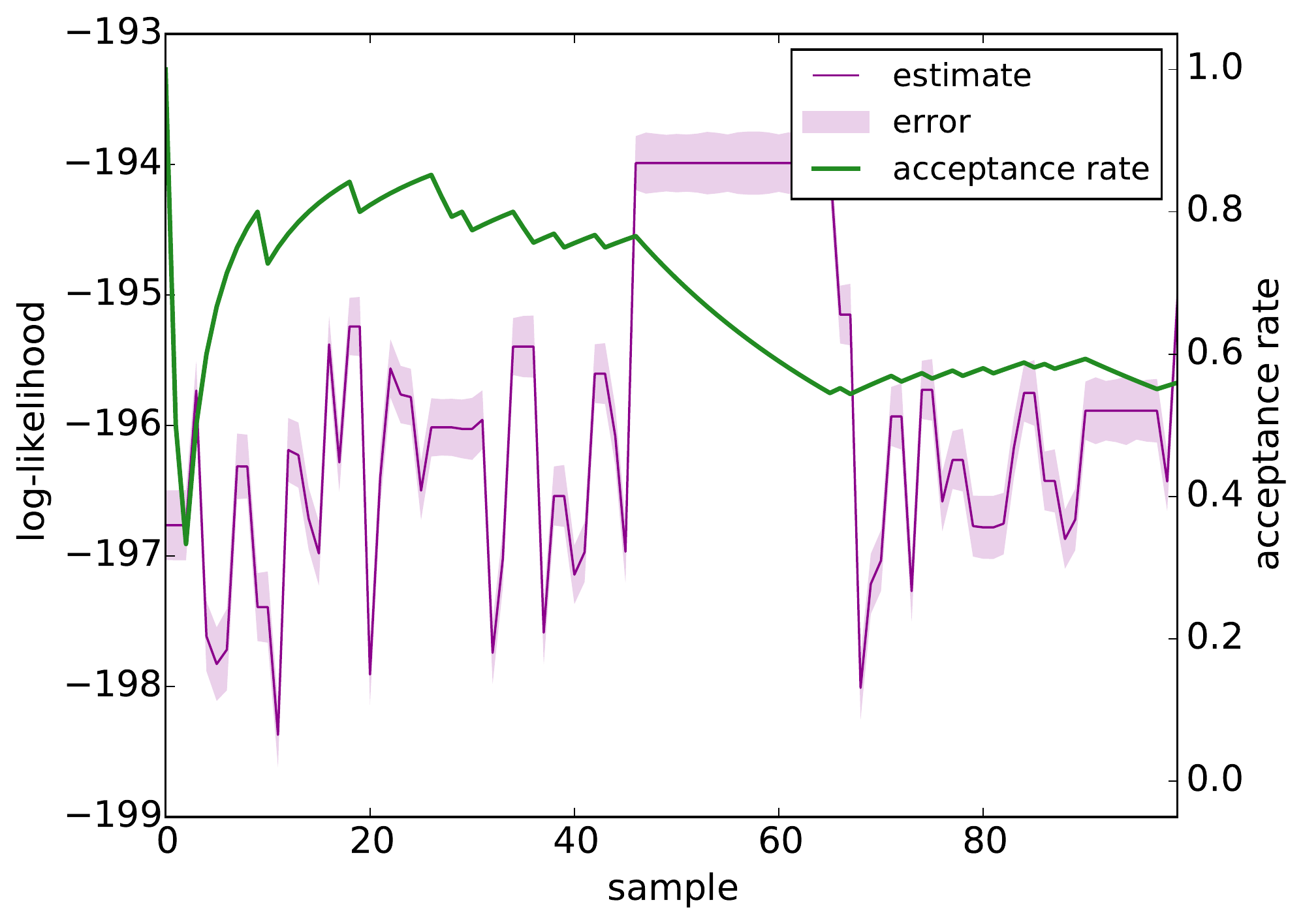}
	\caption{Left: PMCMC posterior samples and estimated posterior distribution.
		Right: estimated marginal likelihoods with corresponding estimated standard deviations
		and rolling MCMC sampling acceptance rate.}
	\label{f:posterior-likelihoods}
\end{figure}

Simulation results are provided in Figure \ref{f:posterior-likelihoods}, where MCMC chain path in parameter space is reported (left plot), together with estimated marginal log-likelihoods for each MCMC sample and rolling MCMC sampling acceptance rate (right plot). The associated standard deviations of marginal log-likelihood are estimated relying on a Taylor-series approximation and are also reported in the right plot of Figure \ref{f:posterior-likelihoods}.

\section{Performance}
\label{s:performance}

We investigate the \PMCMC framework performance for an IBM introduced in section \ref{s:applications}.

\begin{figure}
	\centering
	\includegraphics[width=0.49\textwidth]{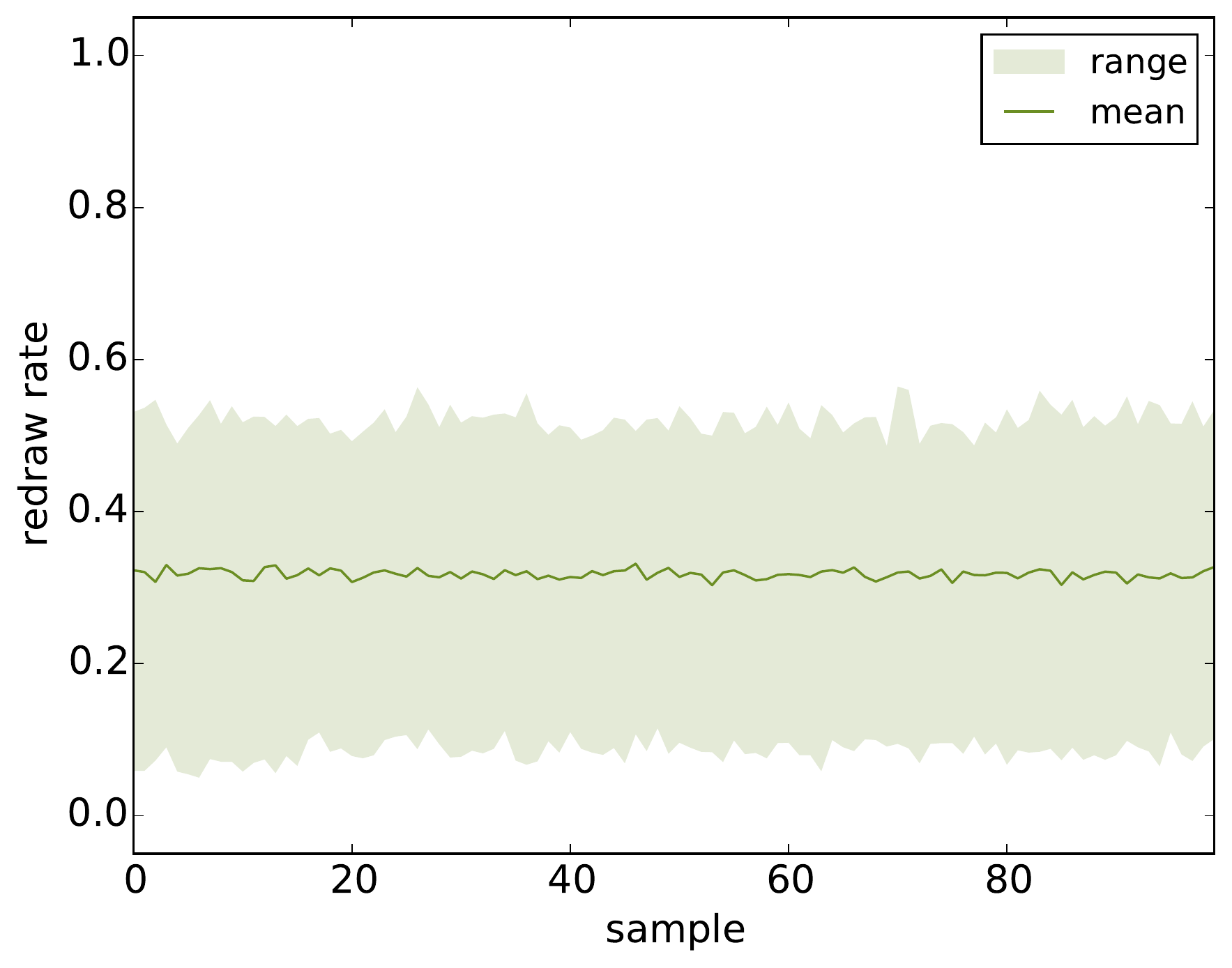}
	\includegraphics[width=0.49\textwidth]{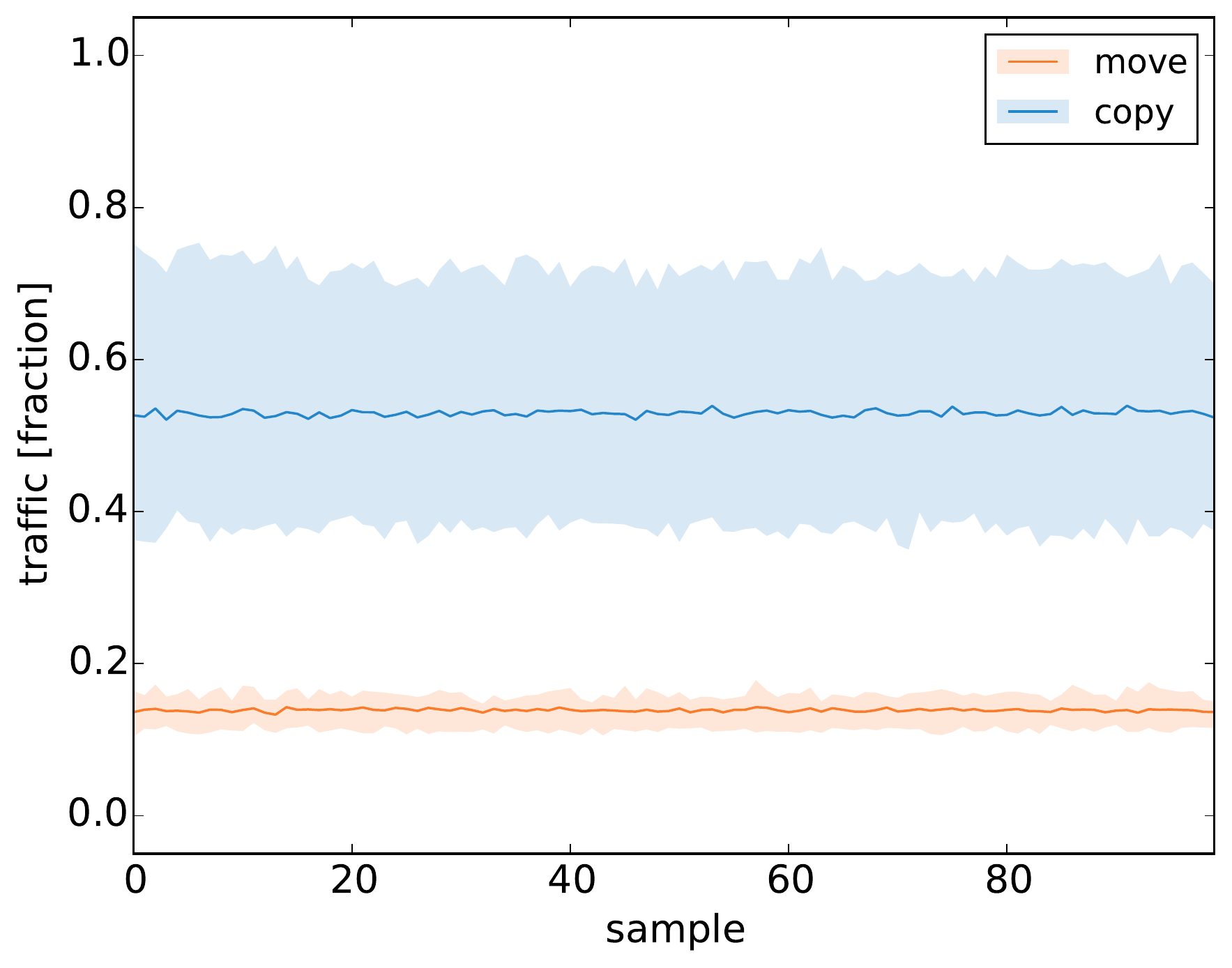}
	\caption{Left: low particle redraw rate ($\approx30\%$)
		is observed throughout all MCMC samples.
		Right: low particle redraw rate causes high local replication (copy) rate $(\approx55\%$); however,
		a low particle move (communication) rate ($\approx15\%$) is maintained due to move-before-replication balancing strategy.
		For each sample, multiple measurements of redraw rate and traffic are obtained at each observation time.
		Solid lines represent average of measured values,
		whereas corresponding surrounding brighter regions represent the range.}
	\label{f:redraw-traffic}
\end{figure}

In Figure \ref{f:redraw-traffic}, particle redraw rates (fraction of all particles that are not removed during particle re-sampling stage in filtering algorithm)
and particle move and copy traffics (fractions of particles that are moved or replicated during the subsequent routing for load balancing) are reported for each MCMC sample.
Multiple measurements of redraw rate and traffic are obtained at each observation time,
average and ranges of which are investigated.
Low particle redraw rate ($\approx30\%$) is observed throughout all MCMC samples due to heterogeneous observation likelihoods obtained from all particles, which lead to a subset of particles that are heavily replicated and dominate the re-sampled ensemble (copy traffic amounts to $(\approx55\%$)). In general, lower redraw rates are expected for models with stronger stochasticity and unstable behavior, both as potential causes for heterogeneous observation likelihoods $\Pobs$.
Despite low particle redraw rate, particle move (communication) rate ($\approx15\%$) is kept low due to move-before-replication balancing strategy, ensuring a minimal communication overhead.

\begin{figure}
	\centering
	\includegraphics[height=0.225\textheight]{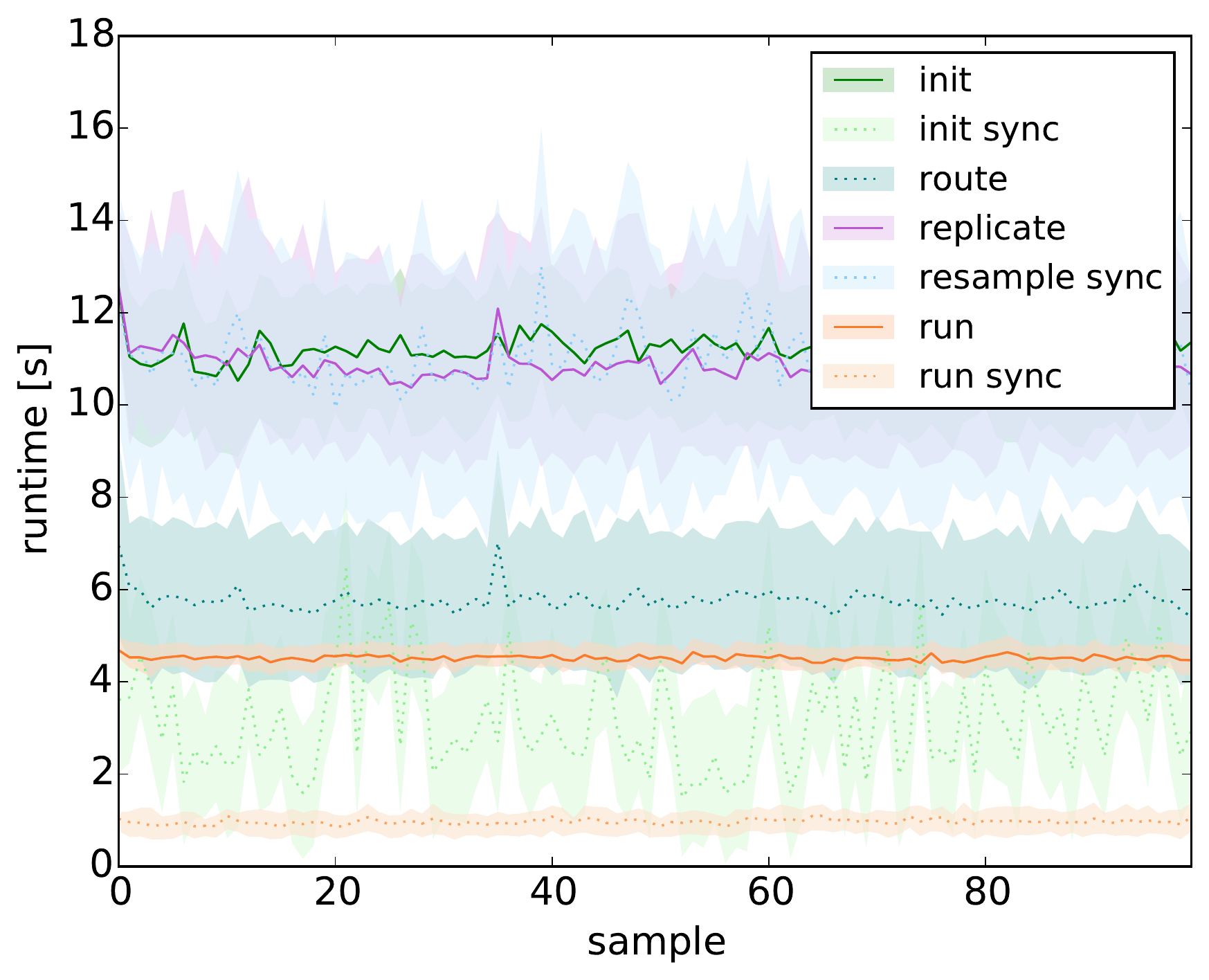}
	\includegraphics[height=0.22\textheight]{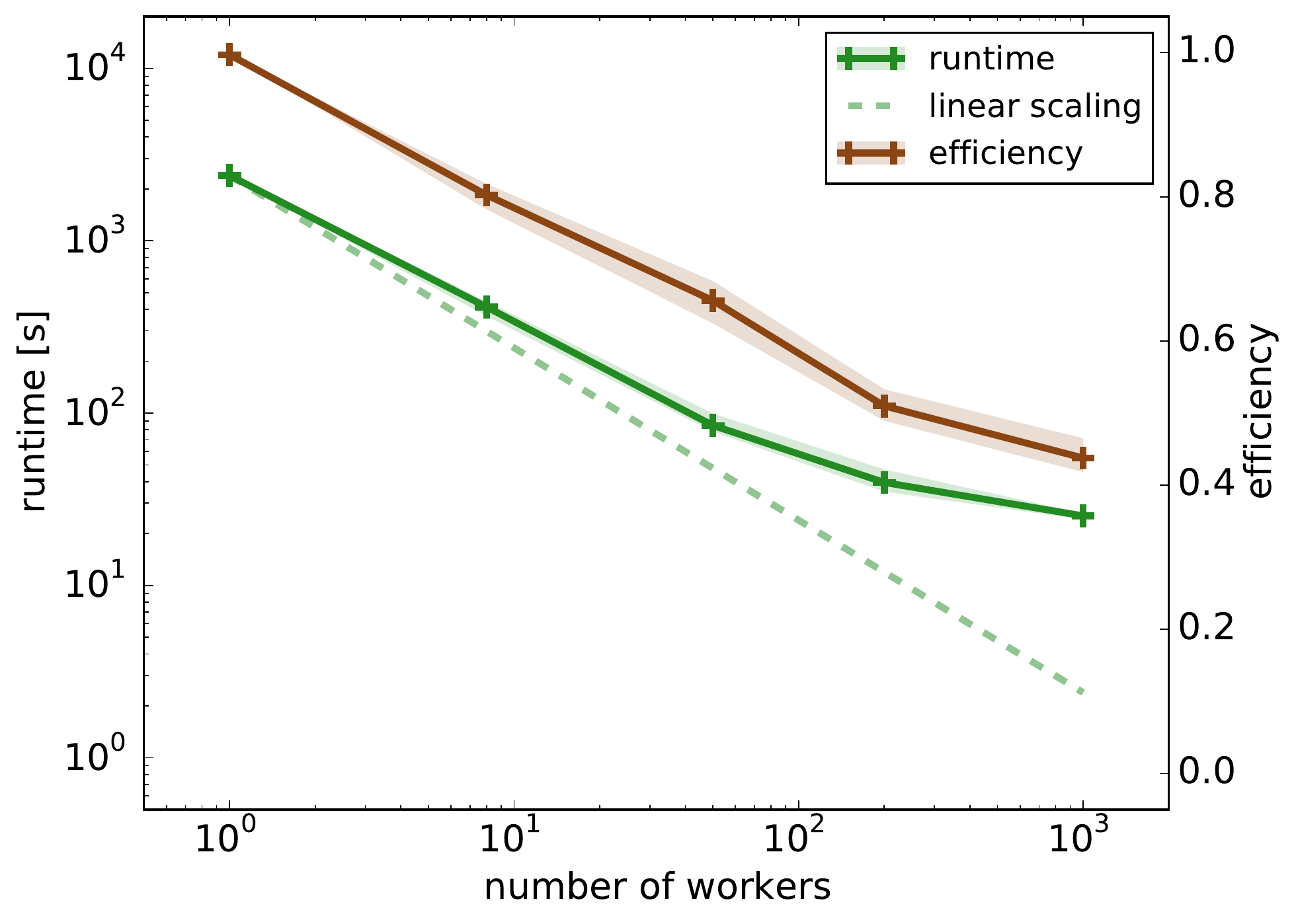}
	\caption{Left: Estimated runtimes for different algorithm simulation and communication stages.
		Only routines with runtimes larger than 2\% of total runtime are depicted.
		For each sample, multiple runtime measurements are obtained
		for each parallel worker;
		dark lines represent measurement averages,
		whereas corresponding surrounding brighter regions represent the 10\% - 90\% percentiles.
		Continuous lines indicate simulation-related routines and dotted lines - communication and synchronization overheads.
		Large variation in initialization \texttt{'init'} runtimes across parallel workers causes significant synchronization overhead \texttt{'init sync'}.
		Particle routing communication routine \texttt{'route'} is dominated by the local copying procedure \texttt{'replicate'} and does not constitute a bottleneck.
		Right: Scalability and efficiency of \PMCMC implementation for parallel Particle MCMC.
		For each simulation, multiple runtime and efficiency measurements are obtained for each MCMC sample;
		dark lines represent measurement averages,
		whereas corresponding surrounding brighter regions represent the 10\% - 90\% percentiles.
		Adaptive particle re-balancing across ensembles
		maintains a relatively high parallelization efficiency up to 200 workers, achieving a significant 100x speed-up compared to a serial code.}
	\label{f:runtimes-scaling}
\end{figure}

Runtimes for different execution stages of all parallel worker processes are aggregated in the left plot of Figure \ref{f:runtimes-scaling}.
Multiple runtime measurements are obtained for each parallel worker at every MCMC sample.
Averages and 10\% - 90\% percentiles of all measured runtimes are investigated.
Both, simulation-related routines as well as communication and synchronization overheads are taken into account.
Large variation in initialization \texttt{'init'} runtimes across parallel workers causes significant synchronization overhead, as can be observed in \texttt{'init sync'}.
Particle routing communication routine \texttt{'route'} is dominated by the local particle copying procedure \texttt{'replicate'} and hence introduces only a relatively small parallelization overhead.
Parallel scalability of \PMCMC for an increasing number of workers (1, 8, 50, 200, 1000) is reported in the right plot of Figure \ref{f:runtimes-scaling}.
Additionally, the resulting parallelization efficiency is provided, which (for each MCMC sample) measures runtime fraction spent on main computing routines such as initialize, run, observe, likelihood, resample, and replicate.

Adaptive particle re-balancing across ensembles
maintains a relatively high parallelization efficiency of $\approx 50\%$ up to 200 parallel workers, despite the challenging complexity of \PMCMC execution graph.
A significant 100x speed-up is achieved compared to a serial code:
2000 particles are simulated in 30 seconds instead of one hour.
In a limiting case of fine-grained parallelization with 1000 workers, where only 2 particles per worker are available,
we observe significant scalability deterioration (compared to theoretical linear scaling) 
due to insufficient amount of computational work (for current IBM) to compensate for the communication overhead.
We note, that multiple independent or mixed MCMC chains can be executed in parallel,
allowing further scalability.

\section{Summary and Outlook}
\label{s:summary}

We presented \PMCMC - a new scalable Python framework for Bayesian inference using PMCMC in stochastic models.
Several orders of magnitude speed-up (100x) of the particle filtering marginal likelihood estimate algorithm were achieved, essentially limited only by the number of particles.
Depending on the available resources, accelerated parallel PMCMC code is capable of reducing the required simulation time for the entire Bayesian inference of predator-prey IBM from almost one month to only several hours, enabling rapid prototyping and calibration of complex IBMs.
The algorithm is expected to scale efficiently on even larger computing clusters,
providing a scalable framework for Bayesian inference in general hidden Markov stochastic models,
accelerating novel insights in data-driven ecological and environmental research.

\section{Acknowledgments}
J{\v S} would like to thank Siddhartha Mishra (SAM, ETH Zurich) for access to EULER, and Peter Reichert (SIAM, Eawag) for fruitful discussions regarding parallelization issues.

\bibliographystyle{unsrt}
\bibliography{library}

%
%
%
%
%

\end{document}